\documentclass[review]{elsarticle}

\usepackage{lineno,hyperref}
\usepackage{upgreek}
\usepackage{units}
\usepackage{multirow}
\usepackage{amssymb}	
\usepackage{amsmath}	

\begin{document}

\begin{frontmatter}

\title{Alternative glues for the production of ATLAS silicon strip modules for the Phase-II upgrade of the ATLAS Inner Detector}

\author[1,2]{Luise Poley}
\author[1]{Ingo Bloch}
\author[3]{Sam Edwards}
\author[1]{Conrad Friedrich}
\author[1]{Ingrid-Maria Gregor}
\author[4]{Tim Jones}
\author[2]{Heiko Lacker}
\author[3]{Simon Pyatt}
\author[2]{Laura Rehnisch}
\author[2]{Dennis Sperlich}
\author[3]{John Wilson}
\address[1]{Deutsches Elektronen Synchrotron, 22607 Hamburg and 15738 Zeuthen}
\address[2]{Humboldt Universit\"{a}t zu Berlin, 12489 Berlin}
\address[3]{University of Birmingham, B15 2TT Birmingham}
\address[4]{University of Liverpool, L69 7ZE Liverpool}

\begin{abstract}
The Phase-II upgrade of the ATLAS detector for the High Luminosity Large Hadron Collider (HL-LHC) includes the replacement of the current Inner Detector with an all-silicon tracker consisting of pixel and strip detectors. The current Phase-II detector layout requires the construction of 20,000 strip detector modules consisting of sensor, circuit boards and readout chips, which are connected mechanically using adhesives. The adhesive used initially between readout chips and circuit board is a silver epoxy glue as was used in the current ATLAS SemiConductor Tracker (SCT). However, this glue has several disadvantages, which motivated the search for an alternative.

This paper presents a study of six ultra-violet (UV) cure glues and a glue pad for possible use in the assembly of silicon strip detector modules for the ATLAS upgrade. 

Trials were carried out to determine the ease of use, thermal conduction and shear strength. Samples were thermally cycled, radiation hardness and corrosion resistance were also determined. These investigations led to the exclusion of three UV cure glues as well as the glue pad.

Three UV cure glues were found to be possible better alternatives than silver loaded glue. Results from electrical tests of first prototype modules constructed using these glues are presented.
\end{abstract}


\end{frontmatter}

\section{Introduction}

Plans for the Large Hadron Collider (LHC) include an upgrade to be completed around 2025 reaching a luminosity of $\mathcal{L} = \unit[6\cdot 10^{34}]{\text{cm}^{-2}\text{s}^{-1}}$ compared to a nominal luminosity of $\mathcal{L} = \unit[1\cdot 10^{34}]{\text{cm}^{-2}\text{s}^{-1}}$, reached in 2011. 
The LHC experiments (ALICE~\cite{ALICELOI}, ATLAS~\cite{ATLASLOI}, CMS~\cite{CMSLOI} and LHCb~\cite{LHCBLOI}) also need to be upgraded for the high luminosity phase in order to be able to cope with an increased primary and secondary interaction rate. For the ATLAS detector, the construction of a new tracking detector is foreseen, since the current Inner Detector (consisting of a pixel detector, a strip-based SemiConductor Tracker (SCT) and a Transition Radiation Tracker (TRT)) is not suited for the anticipated high track density and radiation levels. In the current upgrade plans, the tracker will consist only of a pixel tracker and a strip tracker, arranged in a central region, where silicon sensors are aligned parallel to the beam axis (barrel), and a forward region, where sensors are aligned perpendicular to the beam axis (end-cap). For the silicon strip tracker, the current design foresees about 11,000 modules in the central region and about 8,000 modules in the forward region. Each module is composed of a silicon sensor, one or more circuit boards (hybrids) and readout chips (application-specific integrated circuits (ASICs)). 

Using a silver-loaded epoxy glue, ASICs are glued on to a hybrid, which, later in the module production process, is glued directly on to a silicon sensor using a non-conductive epoxy glue. Electrical connections between the components are made via ultrasonic wire bonds.
In this paper the focus lies on the glue used to connect ASICs and hybrid. The silver-loaded epoxy glue (TRA-DUCT\textregistered 2902~\cite{2902}) contains $\geq \unit[70]{\%}$ (by mass) silver. The high silver content leads to several disadvantages compared to a non-loaded epoxy glue: a high activation by irradiation, a short radiation length $X_0$ and possible corrosion of components consisting of less noble materials. In addition, the glue requires a minimum curing time of six hours, which leads to long construction times for each module.
The silver-loaded epoxy glue was chosen in an early phase of the module design, when ASICs had to be electrically grounded via their backplanes. Since the ASICs in the current design layout are connected to ground by wire bond connections to the hybrid, a conductive glue is no longer required. Therefore the possibility of replacing the silver epoxy glue with a non-conducting adhesive was investigated and is reported in the following.

\section{Selection of alternative adhesives}

To be accepted as a possible replacement a glue should not have any of the disadvantages of the silver epoxy glue. Thus it is required to have a short curing time, a large radiation length and to show neither a high activation after irradiation nor corrosive effects on other components.

In addition, the replacement glue should exhibit a similar or better performance than for silver epoxy glue for the construction and operation of modules. The glue is required to be able to attach ASICs to hybrids (i.e. to attach silicon to gold) with sufficient strength to withstand forces up to $\unit[3.9\cdot10^{-4}]{\text{N}}$ during operation and up to $\unit[4.7\cdot10^{-3}]{\text{N}}$ during transport, to have a low toxicity classification, to be easily dispensed and to show flexibility after curing. As well as a sufficiently strong connection, low thermal impedance and reasonably low relative coefficients of thermal expansion between the components, the operation of modules in the ATLAS detector requires a working temperature range (including shocks) of \unit[-45]{$^\circ$C} to \unit[+80]{$^\circ$C} and a high radiation tolerance were considered mandatory.

\begin{figure}
\includegraphics[width=\textwidth]{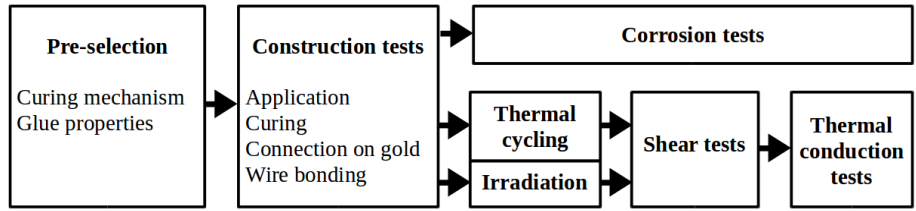}
\caption{Overview of the tests performed with selected glues in chronological order: after glues (UV cure and glue pad) had matched the preselection criteria and were found suitable for the construction of hybrids, prototypes were built and then subjected to tests of the performance required for the future ATLAS Tracker. Possible corrosive effects on aluminium were investigated in parallel.}
\label{fig:00}
\end{figure}

Candidates were selected by searching for commercially available glues matching the specified criteria. First, all selected glues were tested for suitability for the construction process and prototype hybrids, assembled with the replacement candidates, were produced. The prototypes were then subjected to irradiation and thermal cycling. Afterwards the performance of the glues (in particular, their thermal conduction and shear strength) was investigated. A long-term study of possible corrosive effects on aluminium was conducted in parallel to the tests performed with prototypes.
An overview of the different stages of this study is shown in Figure~\ref{fig:00}.

\subsection{Curing mechanisms}

Four glue types were considered during the first selection process:
\begin{enumerate}
\item multi-component adhesives (where mixing two or more adhesive components starts a chemical reaction that leads to a curing of the glue)
\item UV cure adhesives (where the glue is cured by applying UV light)
\item heat curing adhesives (where a curing process is started by heating the glue)
\item pressure sensitive tape (an adhesive film which binds together two components after pressure has been applied)
\end{enumerate}
Adhesives whose curing is started by mixing two or more components were not selected because their curing times were either too long for an efficient production stream or too short for the glue to be dispensed effectively..

UV cure glues, which are cured by applying UV or blue light, are rarely used in combination with gold and have a low thermal conductivity, which limited the number of glues that were available. Six UV cure glues were selected for further investigations.

Heat cure glues, where heating activates the curing process, usually require temperatures of $\mathcal{O}$(\unit[100]{$^\circ$C}), which exceed the components' heat tolerance. Moreover, increased temperatures can lead to deformations of the positioning tools which would complicate the adjustment of glue thicknesses. These glues were therefore not further investigated.

Adhesive films are usually available in thicknesses of several \unit[100]{$\upmu$m}, which exceed the intended glue thickness of \unit[80]{$\upmu$m} between ASIC and hybrid. Despite this, one adhesive film (thickness \unit[300]{$\upmu$m}) was selected as a test of principle.

\subsection{Radiation lengths of silver loaded and UV cure adhesives}

The radiation lengths of different unloaded adhesives were estimated to be about \unit[40]{cm} or \unit[$\approx\,40$]{$\text{g}\cdot\text{cm}^{-2}$} (for a density of \unit[1]{g$\cdot$cm$^{-3}$}).

For silver epoxy glue with a silver content of \unit[70-90]{\%} the radiation length $X_0$ was estimated by:
\begin{equation}
\frac{1}{X_0} = \sum_{i} \frac{f_i}{X_{0,i}},
\end{equation}
(where $f_i$ are the mass fractions of different components of a material and $X_{0,i}$ their radiation lengths) to be between $X_0 = \unit[1.22]{\text{cm}}$ for \unit[70]{\%} silver and $X_0 = \unit[0.95]{\text{cm}}$ for \unit[90]{\%} silver.

This corresponds to a radiation length a factor 30-40 smaller than for an unloaded glue. In the current detector layout, a particle will pass through approximately one glue layer between ASIC and hybrid, when traversing the future silicon strip detector, which corresponds to about \unit[100]{$\upmu$m}, equivalent to \unit[1]{\%} of the silver epoxy glue radiation length of about \unit[1]{cm}.

\subsection{Selected candidates}

A summary of the glues selected for further tests is shown in Table~\ref{tab:01}.
\begin{table}[tp]
\begin{tabular}{cc|ccc}
 & & Cure type & Viscosity & Density \\ 
 & & & $[$mPa$\cdot$s$]$ & $[$g$\cdot$cm$^{-3}]$ \\
\hline
TRA-DUCT\textregistered & 2902~\cite{2902} & silver epoxy & 20000 & 3.2 \\
\hline
\multirow{3}{*}{DYMAX\textregistered} & 3013~\cite{3013} & UV & 150 & 1.04-1.07 \\
 & 3025~\cite{3025} & UV & 300 & 1.05 \\
 & 6-621~\cite{6621} & UV & 800 & 1.08 \\
\multirow{2}{*}{LOCTITE\textregistered} & 3504~\cite{3504} & UV & 800-1300 & 1.1 \\
 & 3525~\cite{3525} & UV & 9500 - 21000 & 1.08 \\
POLYTEC\textregistered& UV 2133~\cite{2133} & UV & not specified & 1.78 \\
3M\textregistered & 5590H~\cite{5590} & pad & - & not specified \\
\end{tabular}
\caption{Overview of silver epoxy glue, used on the current SCT modules to connect ASICs with a hybrid, and possible replacements with selected properties}
\label{tab:01}
\end{table}
All DYMAX\textregistered~glues are declared free of halogens (i.e. $\leq \unit[500]{\text{ppm}}$) and heavy metals by the manufacturer. For the POLYTEC\textregistered~and LOCTITE\textregistered glues no information about the heavy metal content could be obtained.

\section{Construction of module components with alternative glues}

The initial construction method was developed using the silver loaded glue (TRA-DUCT 2902) between ASIC and hybrid. A first series of tests was conducted in order to determine if a glue's mechanical properties were suitable for hybrid construction.

In order to be glued on to a hybrid, groups of ASICs are picked up using a vacuum tool. A thin metal sheet stencil is placed on to the back sides of the ASICs and through precision openings in the stencil, glue is applied. After removing the stencil, each ASIC carries a glue volume of \unit[1.9]{mm$^3$} consisting of five glue dots with heights of \unit[120]{$\upmu$m}.

The ASICs are then positioned above a hybrid at a defined distance of between \unit[60 and 80]{$\upmu$m}, which leads to the \unit[120]{$\upmu$m} glue layer thickness being compressed by \unit[33 to 50]{\%}, so that the five dot pattern forms an effective connection between the components. The ASICs are held on the vacuum tool at a fixed height above the hybrid until the glue is cured.

After gluing, ASICs and hybrid are connected electrically by aluminium wire bonds (wire bonding step).

\subsection{Glue dispensing}

The standard stencil was designed for Tra-Duct~2902~\textregistered, which has a high viscosity of \unit[20,000]{mPa$\cdot$s}. The UV cure glues under investigation have different viscosities ranging from highly viscous to highly fluid (see Table~\ref{tab:01}). In an initial series of tests it was found that only one candidate (POLYTEC\textregistered~UV 2133, with a high viscosity) could be used with the glue stencil. Highly fluid glues (DYMAX\textregistered~3013, 3025 and 6-621) led to the glue spreading outside the five dot pattern into the gap between ASIC and stencil. Glues of a honey-like consistency (LOCTITE 3504 and 3525) led to several ASICs being glued to the stencil.

The investigation of possible alternatives to replace the stencil resulted in two options:
\begin{itemize}
\item a microlitre pipette, which allows a predefined amount of glue to be dispensed manually
\item an automatic glue dispenser, where glue is dispensed by applying pressure for a specified amount of time to drive the plunger of a glue syringe
\end{itemize}
The method used must be precise enough that the glue thickness is between \unit[60]{$\upmu$m} and \unit[80]{$\upmu$m}, that the glue covers a sufficient area of the ASIC but does not squeeze out from underneath the chip.

An estimation of the required volume precision can be made by requiring that the glue, at a thickness of between \unit[60]{$\upmu$m} and \unit[80]{$\upmu$m}, covers a sufficient area under an ASIC (\unit[$7.7\cdot7.9$]{mm$^2$}), but does not squeeze out from underneath the chip.

\begin{figure}
\includegraphics[width=\textwidth]{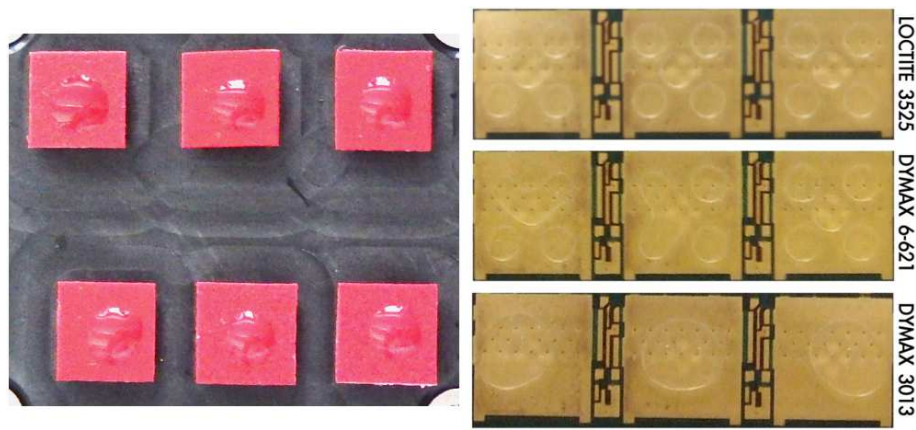}
\caption{UV cure glues dispensed with a microlitre pipette on red plastic (left, DYMAX\textregistered~6-621) and a fully automated glue dispenser (right). While a pipette allows only the application of a single glue spot, an automated glue dispenser can reproduce the original five-dot glue pattern for DYMAX\textregistered~6-621 and LOCTITE\textregistered~3525.}
\label{fig:01}
\end{figure}

While a microlitre pipette showed a sufficient volume precision for highly fluid glues, positioning was done manually and only one-dot glue patterns could be achieved. Although the microlitre pipette was used for the construction of prototypes, alternatives for dispensing the glue were investigated. Automatic glue dispensers used at the Universities of Birmingham and Glasgow were found to be able to produce dot patterns with high precision alignment for glues with different viscosities (see Figure~\ref{fig:01}) with good testing results.

Both dispensing methods were found to produce good prototypes with no glue squeezing out from below the ASICs. Also the glue dispenser and the microlitre pipette were found to show smaller variations in the amount of UV cure glue per ASIC than the stencil did with the silver epoxy glue.

\subsection{Curing}

In order to cure the UV glue between ASICs and hybrid, UV light was directed at the \unit[80]{$\upmu$m} glue layer gap using four light guides (each of diameter \unit[2]{mm}) connected to a commercially available mercury arc lamp. It was found that in all trials the glue was completely cured after applying UV light for a total of \unit[200]{s}.

UV LEDs were investigated as a potentially more cost-efficient UV light source. UV LEDs~\cite{UVLED}, with a \unit[1]{W} rating and operating at \unit[350]{mA} with an emitted wavelength of \unit[405]{nm}, were chosen to match the curing wavelength (\unit[350-420]{nm}) of the chosen glues~\cite{3025, 6621, 3013}. The UV LEDs were positioned in an aluminium frame with one LED next to each ASIC, with the option to connect them to a cooling unit and to install them on a vacuum jig surrounding a hybrid.

\begin{figure}
\includegraphics[width=\textwidth]{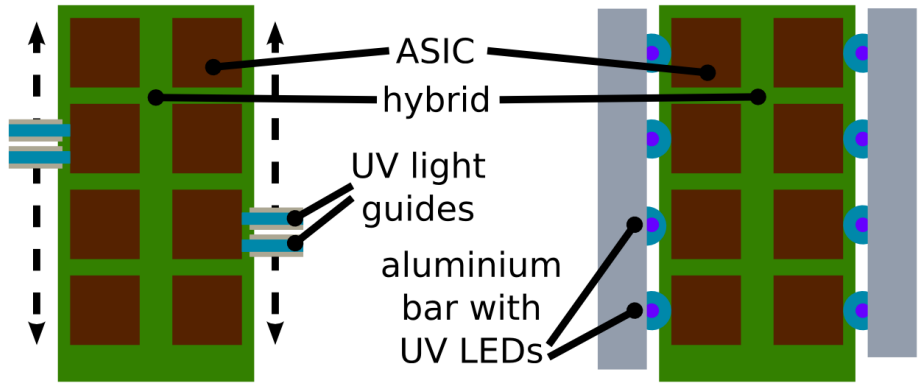}
\caption{Scheme of UV curing setups for gluing ASICs on to a hybrid using light guides connected to a mercury arc lamp (left) and UV LEDs mounted on an aluminium bar (right), not to scale. While the use of a mercury arc lamp requires the light guides to be moved along the edge of the hybrid in order to reach all ASICs (dashed lines), UV LEDs could be used in a static setup where one UV LED was positioned next to the centre of each ASIC. In both setups, UV light is applied from the only accessible edge of the ASICs i.e. from the outer edges of the hybrid.}
\label{pic:UVsetups}
\end{figure}

Figure~\ref{pic:UVsetups} shows UV curing setups using UV LEDs and light guides connected to a mercury arc lamp in comparison.

For both setups, a sufficient curing of the whole glue area under each ASIC was confirmed by disassembling the glued components after curing and inspecting the glue layers visually or by trying to remove uncured glue with acetone.
\begin{center}
\begin{figure}
\includegraphics[width=0.5\textwidth]{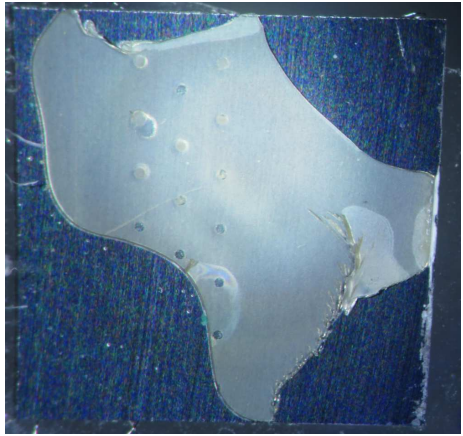}
\caption{Backside of an ASIC glued with LOCTITE\textregistered~3504, removed after curing with UV light guides connected to a mercury arc lamp. The glue layer is broken around three corners of the ASIC and shows additional cracks toward the ASIC centre. No regions of uncured glue were found.}
\label{pic:cured}
\end{figure}
\end{center}
Figure~\ref{pic:cured} shows an example of an ASIC glued with LOCTITE\textregistered~3504, removed from a hybrid in a shear test (see section~\ref{subsec:ss}).

The UV LED curing setup was found to cure all glue layers between ASICs and hybrid completely within \unit[120]{s}, sufficiently fast also for mass production. The method is more easily applicable than the use of UV lamp and light guides.

\subsection{ASIC wire bonding}

It was found that, after curing, the UV glue layers were still moderately elastic, so that bending the hybrid slightly did not loosen the ASICs glued to it.

This flexibility is a useful feature of UV cure glues. However, a glue layer of high elasticity could potentially cause problems during the wire bonding step as rigid surfaces are required to ensure a sufficiently strong wire bond connection and to allow the wire bonding machine to operate at maximum speed.

All UV cure glues and the glue pad under investigation were found to provide good wire bond connections between ASIC and hybrid. Hybrids glued with UV cure glue (LOCTITE\textregistered~3525, DYMAX\textregistered~3013 and 6-621) showed good thermal and electrical performance when powered and connected to a data readout system. Results were comparable to those of a hybrid glued with sliver epoxy glue.

\subsection{Conclusion of construction tests}

After testing glue application, curing and wire bonding, all glues were found to be suitable for the construction of hybrids. Except for the method of glue application and curing, using UV cure glues as alternatives to Tra-Duct\textregistered~2902 did not require modification of the assembly procedure.

\section{Thermal and mechanical glue properties before and after irradiation and thermal cycling}

\subsection{Coefficients of thermal expansion}

After hybrids are constructed at room temperature, their operational temperature in the detector will be \unit[-20]{$^{\circ}$C}, corresponding to a temperature change of about \unit[40]{K}, which can induce stress in components with different coefficients of thermal expansion. Table~\ref{tab:CTE} summarises the coefficients of thermal expansion (CTE) for several UV cure glues under investigation in comparison with module component constituents.
\begin{table}[tp]
\begin{tabular}{c|c|c}
\multirow{2}{*}{Material} & \multirow{2}{*}{Component} & CTE (linear), \\
 & & $[10^{-6}$ 1/K$]$ \\
\hline
Silicon & ASIC & 2.6 \\
Gold & Hybrid & 14.2 \\
Polyimide & Hybrid (coating) & $\approx20$~\cite{38492} \\
Copper & Hybrid & 16.5 \\
\hline
TRA-DUCT\textregistered~2902 & silver epoxy glue & 49.0~\cite{2902} \\
DYMAX\textregistered~3013 & UV cure glue & not speficied \\
DYMAX\textregistered~3025 & UV cure glue & not speficied \\
DYMAX\textregistered~6-621 & UV cure glue & not speficied \\
LOCTITE\textregistered~3504 & UV cure glue & 80.0~\cite{3504} \\
LOCTITE\textregistered~3525 & UV cure glue & 97.0~\cite{3525} \\
POLYTEC\textregistered~UV 2133 & UV cure glue & 35.0~\cite{2133} \\
3M\textregistered~5590H & glue pad & not specified \\
\end{tabular}
\caption{List of materials used in components for silicon strip detector modules with coefficients of thermal expansion (CTE). The adhesives show significantly higher CTE than constituents of components. DYMAX\textregistered~UV cure glues with no specified CTE can be assumed to have values of CTE comparable to LOCTITE\textregistered~UV cure glue because of their similar viscosity, density and composition.}
\label{tab:CTE}
\end{table}
The different CTE could, over the lifetime of a hybrid, lead to different possible failure scenarios: large mechanical stresses could lead to damaged ASICs or hybrids or failing glue joints, which would result in low shear strengths or a possible overheating of ASICs.
In order to evaluate these effects on glue joints, different tests were performed:
\begin{itemize}
 \item single ASIC to hybrid glue joints were shear tested after irradiation and thermal cycling (see section~\ref{subsec:ss})
 \item full hybrids were populated, partially irradiated and their thermal performances monitored during operation (see section~\ref{subsec:tc})
\end{itemize}
No adverse effects originating from the mechanical stress induced by the different CTE values could be determined. Based on these observations the less positive CTE range of the UV cured adhesives was not considered a criterion for exclusion.

\subsection{Impact of thermal cycling on glue connections}

The possible long-term weakening of glue joints, caused by repeated temperature changes, was investigated in a climate chamber. Each hybrid was subjected to the same thermal cycling tests as the ATLAS upgrade prototype sensors~\cite{nobu}, 

100 cycles of \unit[14]{hrs} length per cycle were performed with temperature varying between \unit[-20]{$^\circ$C} and \unit[+50]{$^\circ$C} over \unit[60]{min} and in a controlled low relative humidity (\unit[$\leq15$]{\%}).

One hybrid, glued with silver epoxy glue, and two hybrids, each glued with UV cure glue and glue pads, were subjected to this thermal cycling. Of these latter two, one was constructed using glass dummy ASICs to allow for optical inspection of the glue joints. Each hybrid was populated with 20 ASICs or glass dummy ASICs.

After thermal cycling all glue connections were still intact. Visible changes were observed for one glue (POLYTEC\textregistered~UV 2133). Here the glue dot surface exposed to the air had turned white, indicating a structural change on the surface. No visible changes were observed for any other glue.

Both the hybrid glued with silver epoxy glue and the hybrid glued with either UV cure glues or glue pad, were subjected to shear tests after (presented in section~\ref{subsec:ss}) thermal cycling in order to determine possible impacts of the thermal cycling on the glue joints.

\subsection{Impact of irradiation on glue connections}

Information concerning the impact of irradiation on the performance of the adhesives under study was not provided by the manufacturers. A total fluence of up to \unit[10$^{15}$]{n$_{\text{eq}}$/cm$^{2}$} (\unit[1]{MeV} neutron equivalent), mainly from hadrons, is expected in the future ATLAS strip tracker after a runtime of ten years~\cite{ATLASLOI}. As this fluence is ten times that expected for the current ATLAS SCT, radiation hardness is one of the main requirements for all materials used for construction.

the irradiation corresponded to twice the expected total ATLAS fluence. Irradiaton of polymers, such as glues, can lead to either a better cross-linking of the polymers or a worse interconnection by breaking up large molecules into shorter chains, depending on the irradiation type and energy, material composition and temperature~\cite{rad1}. Irradiation tests were performed with \unit[23]{MeV} protons at the Karlsruhe Kompaktzyklotron (KAZ) at a temperature of \unit[-20]{$^{\circ}$C}.

Three test structures were constructed for irradiation: two hybrids, populated with 20 ASICs each, and a polyethylene plate. One hybrid was glued with silver epoxy while the other was constructed with UV cure glues and a glue pad. Glue spots were also dispensed on to a polyethylene plate in order to check directly for visible changes in the glue.

After irradiation with \unit[$2\cdot10^{15}$]{n$_{\text{eq}}$/cm$^2$} it was found that no ASICs had become detached from their respective hybrids. Visible changes could be observed for all UV cure glues: POLYTEC\textregistered~UV 2133 had turned from light brown to white, the DYMAX\textregistered~and LOCTITE\textregistered~glues, which are transparent with different shades of yellow, had turned to a darker shade of yellow, with the extent of the colour change varying for the different glues.
Although the observed colour changes were difficult to quantify, the thermo-mechanical characteristics of the glue joints (shear strength and thermal conduction) before and after irradiation were compared (see sections~\ref{subsec:ss} and~\ref{subsec:tc}).

\subsection{Activation of test structures after irradiation}

The level of activation of an irradiated hybrid was determined by measuring the spectrum of photons emitted by the sample in the keV to MeV range. Hybrids glued either with silver epoxy glue or with UV cure glues showed similar levels of activation. Comparing the gamma spectra of irradiated hybrids, the one with silver epoxy glue shows an additional peak. Due to the complex composition of the hybrid and the dominant contributions of its metal components, mainly copper, the comparably small activation contribution of the silver (\unit[3-4]{\%} of the overall hybrid mass) could not be identified.

The measurements suggest that the use of unloaded glues does lead to lower levels of activation, but not significantly.

\subsection{Shear strength}
\label{subsec:ss}

In addition to mechanical stress caused by contraction and expansion due to temperature changes, gravity and acceleration during transport act on glue joints. For an ASIC with a weight of \unit[0.04]{g}, gravity exerts up to \unit[$4.7\cdot10^{-3}$]{N} as shear force or pull force during transport, depending on a hybrid's alignment angle.

Shear tests were performed in order to determine the connection strength of an ASIC on a hybrid glued with a specific adhesive. For the shear test, a hybrid was screwed down on a holding structure and aligned vertically under a movable shear tool (spatula). The tool's position was adjusted manually with microscrews so that the spatula made contact with the upper edge of the ASIC.

After positioning, a steering programme was started which lowered the spatula at a predefined rate of \unit[0.5]{mm/s} and measured the force required to move the spatula at this constant rate against an ASIC. A shear test was stopped when the force dropped by \unit[80]{\%}, which usually occured when an ASIC was removed from a hybrid. The resulting peak force was taken as a quality criterion for a glue joint's shear strength. For the initial series of tests b-grade components were used, i.e. ASICs with small mechanical damages, such as cracks or broken edges, so that applying a shear force in some cases caused an ASIC to splinter or break instead of being removed from a hybrid. In these cases the determined shear force was taken only as a lower estimate for the actual shear strength of a glue joint.

The resulting peak shear forces or highest force reached before an ASIC broke (numbers in parentheses) are shown in Table~\ref{tab:06}.
\begin{table}[tp]
\begin{tabular}{c|cc|cc}
& \multicolumn{4}{|c}{Peak shear force, $[\text{N}]$} \\
& \multicolumn{2}{|c|}{after irradiation} & \multicolumn{2}{|c}{after thermal cycling} \\
\hline
& \multicolumn{2}{|c|}{Hybrid} & \multicolumn{2}{|c}{Hybrid} \\
& I & II & I & II \\
\hline
TRA-DUCT\textregistered~2902 & 68 & 128 & (17) & 104 \\
\hline
DYMAX\textregistered~3013 & 89 & 127 & 68 & - \\
DYMAX\textregistered~3025 & 171 & (146) & 53 & (146) \\
DYMAX\textregistered~6-621 & 124 & (123) & (44) & (72) \\
LOCTITE\textregistered~3504 & 49 & (23) & 118 & (53) \\
LOCTITE\textregistered~3525 & 109 & (136) & 83 & 163 \\
POLYTEC\textregistered~UV 2133 & - & - & 20 & 85 \\
3M\textregistered~5590H & $\leq2$ & $\leq2$ & $\leq2$ & $\leq2$ \\
\end{tabular}
\caption{Peak shear forces for ASICs glued with silver epoxy glue (TRA-DUCT\textregistered 2902) and possible replacement glues, after irradiation and thermal cycles. For each glue, two ASICs were used for shear tests on each hybrid (columns I and II). Numbers in parenthesis represent measurements where an ASIC was damaged instead of being removed. POLYTEC\textregistered~UV 2133 and 3M\textregistered~5590H were found to be insufficient.}
\label{tab:06}
\end{table}
One of the glues (POLYTEC\textregistered~UV 2133) had become brittle after irradiation, so that the glued ASICs fell off the hybrid already during handling. The glue joint was found to have failed inside the glue rather than at the joint between glue and ASIC or hybrid. Thus this glue was rejected as possible replacement for the silver epoxy glue.

In the case of the glue pad, the spatula moved against an ASIC with the preset minimum force of \unit[2]{N}, moving it continuously downwards. As a consequence, the glue pad was rejected as a valid glue alternative, too.

All remaining five UV cure glues under investigation had a sufficiently high shear strength and even exceeded the results for silver epoxy glue connections.

\subsection{Corrosion of aluminium by UV cure glues}

Using a glue which contains a large amount of silver can lead to diffusion between a noble and a less noble metal. This can lead to corrosion of the less noble material, when the material with a higher standard electrode potential, draws electrons from the less noble material.

In order to monitor possible corrosive effects of the glues under investigation, each glue was dispensed on an aluminium foil in a clean room environment. After several weeks, the contact area between one glue (LOCTITE\textregistered~3504) and aluminium showed a colour change from transparent to white. While no actual corrosion was observed, a colour change might be an indicator for a chemical reaction between the two materials, hence this specific glue (LOCTITE\textregistered~3504) was rejected as a possible replacement. No changes were observed for any other glue.

\subsection{Additional considerations}

For the remaining four UV cure glues, additional considerations were taken into account (see Table~\ref{tab:07}):
\begin{table}[tp]
\begin{tabular}{c|c|l}
Glue & Rejected & Additional considerations \\
\hline
DYMAX\textregistered~3013 & - & toxicity: irritant \\
DYMAX\textregistered~3025 & - & toxicity: toxic \\
DYMAX\textregistered~6-621 & - & toxicity: irritant \\
LOCTITE\textregistered~3504 & corrosive effects & \\
LOCTITE\textregistered~3525 & - & toxicity: harmful \\
 & & low glass transition\\
 & & temperature \\
POLYTEC\textregistered~UV 2133 & brittle after irradiation & \\
3M\textregistered~5590H & low shear strength & \\
\end{tabular}
\caption{Overview of test results of seven glues under investigation: three glues were rejected as possible replacements. Among the remaining four UV cure glues DYMAX\textregistered~3013 and 6-621 are considered preferable because of a lower toxicity classification.}
\label{tab:07}
\end{table}
\begin{itemize}
\item Glass transition temperature, i.e. the temperature which marks the transition from solid to the glass-like fluid state. In order to avoid glass transition of a glue layer under an operated ASIC, a glue's glass transition temperature should be high (cf. to \unit[52]{$^\circ$C~\cite{2902}} for silver epoxy glue). Glass transition temperatures were only provided in the data sheets of two glues: \unit[74]{$^\circ$C} for DYMAX\textregistered~6-621~\cite{6621} and \unit[43]{$^\circ$C} for LOCTITE\textregistered~3525~\cite{3525}. Using differential scanning calorimetry~\cite{DSC}, all remaining glue candidates were later confirmed to have glass transition temperatures higher than those attained by hybrids in normal operation.
\item A high toxicity classification could require additional safety measures for handling and would thus complicate the construction process. While DYMAX\textregistered~3013 and DYMAX\textregistered~6-621 are classified as Xi~\cite{3013, 6621} (irritant), DYMAX\textregistered~3025 is classified T~\cite{3025} (toxic) and may require additional inhalatory system protection. LOCTITE\textregistered~3525 is Xn~\cite{3525} (harmful to the environment), which requires additional measures for transport and disposal. Hence DYMAX\textregistered~3013 and DYMAX\textregistered~6-621 should be preferred as final candidates.
\end{itemize}
Finally the glue reworkability, i.e. the possibility to remove ASICs found faulty in electrical tests, was investigated for the remaining three candidates. ASICs glued to a substrate using either UV cure glue or silver epoxy glue were exposed to a stream of air heated up to \unit[200]{$^{\circ}$C} and afterwards removed from the substrate. All UV cure glues led to easily removable ASICs after heating and provided a better reworkability than the silver filled epoxy glue.

\subsection{Thermal performance evaluation}
\label{subsec:tc}

During operation, the heat dissipated in the ASICs is transferred to the cooling structure via a thermal path which includes the glue joint between ASIC and hybrid.

Specific thermal conductivities were provided by the manufacturers for the glue pad (\unit[3.0]{$\frac{\text{W}}{\text{m}\cdot\text{K}}$}) and one of the UV cure glues (\unit[0.1]{$\frac{\text{W}}{\text{m}\cdot\text{K}}$} for LOCTITE\textregistered~3504). By comparison, the specific thermal expansion for silver loaded epoxy glue is \unit[3.0]{$\frac{\text{W}}{\text{m}\cdot\text{K}}$}.

In order to investigate the thermal conductivity of the remaining UV cure glues of interest, functional components were used to construct electrically working hybrids. Using either silver epoxy glue or one of the most promising UV cure glue candidates (DYMAX\textregistered~3013, DYMAX\textregistered~\mbox{6-621} or LOCTITE\textregistered~3525), four hybrids were built under the same conditions. Due to a limited number of available components, hybrids of a different geometry, requiring only twelve ASICs per hybrid, were used for these tests.
Since a major concern of this study was the glue behaviour after irradiation, one half of each hybrid, i.e. six out of twelve readout chips, were irradiated up to \unit[$2\cdot10^{15}$]{n$_{\text{eq}}$/cm$^2$} using \unit[23]{MeV} protons.

\begin{center}
\begin{figure}
\includegraphics[width=0.7\textwidth]{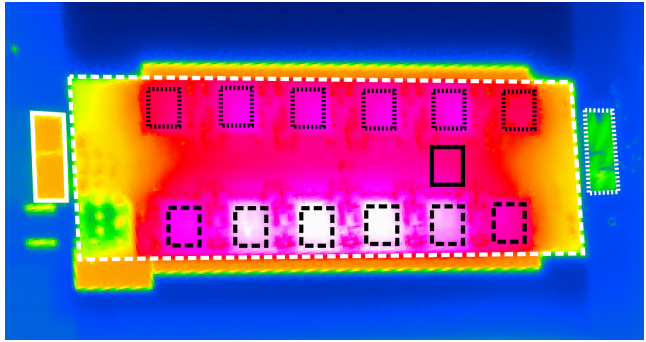}
\caption{Thermal image, taken with a microbolometer based thermal camera, of a silver epoxy glued hybrid in operation. The hybrid (outline indicated with dashed white line) is connected to power (solid white line) and data readout (fine dashed white line) on a testing circuit board (blue area), positioned on a cooling jig. ASIC temperatures were monitored by marking their outlines on the thermal camera image (dashed black lines, fine dashed lines for irradiated ASICs) and using their average, calculated by the thermal camera analysis software. Irradiated ASICs (bottom) were found to reach higher temperatures during operation than unirradiated ASICs. The hybrid temperature was monitored in a similar measurement area between the ASICs (solid black line).}
\label{pic:tc}
\end{figure}
\end{center}

\begin{center}
\begin{figure}
\includegraphics[width=0.8\textwidth]{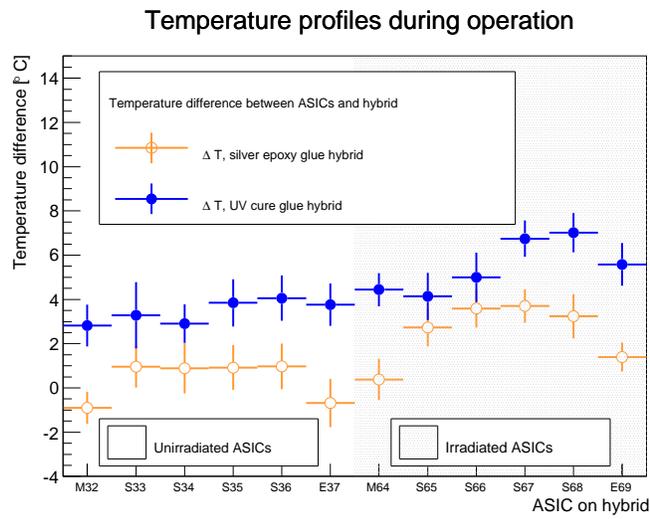}
\caption{Comparison of the temperature profiles measured on a hybrid glued with silver epoxy glue and a hybrid glued with UV cure glue (DYMAX\textregistered~6-621) for unirradiated (white background) and irradiated (grey hatched background) hybrid areas. Temperature differences between ASICs and hybrid were found to be consistently larger for the UV cure glued hybrid, compared to the silver epoxy glued hybrid, by {\unit[2-5]{$^{\circ}$C}}.}
\label{fig:tc2}
\end{figure}
\end{center}

Afterwards, each hybrid was operated on an FR4 circuit board, placed on a cooling jig cooled down to \unit[15]{$^{\circ}$C}.
Temperatures of ASICs and hybrid were estimated from emissions registered using a microbolometer based thermal camera~\cite{vario}. Temperatures were monitored on each ASIC, in comparison with the temperature of the hybrid, the circuit board and the cooling jig, were measured before and during operation of the hybrid (see figure~\ref{pic:tc}).
For each ASIC area, an average temperature was calculated and its minimum and maximum temperatures were used to estimate the temperature uncertainty. Afterwards, the temperature profiles of ASICs glued with UV cure glues were compared to the temperatures observed on a hybrid glued with silver epoxy glue.
Figure~\ref{fig:tc2} shows the comparison for a hybrid glued with DYMAX\textregistered~6-621.
It was found that the overall temperature profile of hybrids glued with UV cure glue is comparable to hybrids glued with silver epoxy glue, both before and after irradiation. The silver epoxy glue showed a smaller temperature difference between hybrid and ASICs than the UV glued hybrids. Although there is a significant temperature variation from ASIC to ASIC, the overall difference for the UV glues is acceptable. This is understandable, as the glue thickness is \unit[60-80]{$\upmu$m} and is therefore only one in many contributions to the thermal path from the active area of the ASIC to the centre of the hybrid.

In summary, all UV cure glues still considered at this point provided a satisfactory thermal connection between ASICs and hybrid which was reasonably close to the thermal connection provided by silver epoxy glue.

\section{Conclusion and Outlook}

Seven glues (six UV cure glues and one glue pad) were investigated, as possible alternatives to the silver epoxy glue currently in use, to connect ASICs and hybrids for silicon strip modules in the ATLAS detector. All glues were tested and found to be suitable for the construction of hybrids and to provide sufficient thermal conduction between the components.
Two glues (POLYTEC\textregistered~UV 2133 and 3M\textregistered~5590H) were rejected during shear tests after irradiation, where their shear strength was found to be insufficient. Five glues still provided sufficient shear strength after thermal cycles and irradiation.
One of these glues (LOCTITE\textregistered~3504) showed indications of a chemical reaction with aluminium and was subsequently rejected. Among the remaining four candidates no final choice has been made, but two glues (DYMAX\textregistered~3013 and 6-621) are preferred over the others.  DYMAX\textregistered~3025 and LOCTITE\textregistered~3525 have a higher toxicity classification and LOCTITE\textregistered~3525 also has a low glass transition temperature. Thermal tests have been conducted for the three preferred UV cure glues and no significant difference has been found in the thermal behaviour of hybrids glued with silver epoxy glue or UV cure glue.

All of the remaining UV cure glues were found to provide all the characteristics required for hybrid construction, without having the disadvantages of a metal-filled epoxy glue.

First hybrids, glued with UV cure glues DYMAX\textregistered~3013, DYMAX\textregistered~6-621 and LOCTITE\textregistered~3525, showed electrical performances comparable to a hybrid glued with silver epoxy glue.

Before a conclusive decision on the use of UV cure glue in module production is made, module prototypes will be constructed by using hybrids to which ASICs were connected with UV cure glues.

In a further step, UV cure glues will be evaluated as well for the glue connection between hybrid and sensor, where their short curing time and flexibility after curing would also provide an advantage compared to the non-conductive epoxy glue (FH 5313) currently in use~\cite{epolite}.

\section{Acknowledgements}

This work was supported by the Helmholtz-Alliance ``Physics at the Terascale`` project ``Enabling Technologies for Silicon Microstrip Tracking Detectors at the HL-LHC''.

The authors would like to thank Dr. Alexander Dierlamm and Felix B\"{o}gelspacher for their help with the planning and realisation of the irradiations conducted for this study at Karlsruhe Kompaktzyklotron (KAZ).

Additional thanks goes to the ATLAS strip tracker community, especially Dr. Tony Affolder, for providing support and advice for the conducted study.

\section*{References}

\bibliography{bib.bib}

\end{document}